# Comment on 'Superhard Semiconducting Optically Transparent High Pressure Phase of Boron'

A recent Letter by Zarechnaya et al. [1] examined the crystal structure, chemical bonding and hardness of the densest and hardest known phase of boron, γ-$B_{28}$ [2,3,4]. In this Comment we wish to point out several important issues with this Letter. A central claim for novelty of [1] is the finding that γ-$B_{28}$ possesses an unusual set of properties, being a "wide band gap semiconductor, superhard, optically transparent". γ-$B_{28}$ was probably first synthesized in 1965 [6], but only recently established as a pure boron phase with a wide stability field on the phase diagram [2]. Its structure was solved [2] and confirmed by subsequent studies [4,1]. Its superhardness, shown in [3], was confirmed in [1]. Moreover, all the 'novel' properties are also possessed by α-boron and to a large extent by β-boron – the two most usual allotropes of boron. Vickers hardness of γ-$B_{28}$ is 50-58 GPa [3,1], slightly harder than β-boron (45 GPa – see [3] and references therein), for which Zarechnaya et al. [1] only quoted anomalously low values (25-30 GPa). The claim that γ-$B_{28}$ is the second hardest elemental material after crystalline diamond, ignores recent works [11-14].

Zarechnaya et al. state that γ-$B_{28}$ has density 2.54 g/cm$^3$, which is "about 1% higher than densities of known (α-, β-, "tetragonal") boron modifications". The density contrasts are in fact much larger [7]: 8.3% with β-boron, 6.8% with T-192 phase, and 3.2% with α-boron. The explanation [1] that γ-$B_{28}$ is densest because only this phase contains additional boron atoms between the $B_{12}$-icosahedra [1] ignores the fact that structures of both β-boron and T-192 phases contain a large number of non-icosahedral atoms [7]. Statement that "the existence of ''tetragonal boron'' as a modification of pure elemental boron or as a boron-rich nitride or carbide has been a subject of controversy" mixes up the two known and very different tetragonal phases: T-50 (proven [9] to be a compound) and T-192 (pure elemental phase [9]).

Zarechnaya et al.'s two estimates of the density of γ-$B_{28}$ differ by 1%: 2.52 g/cm$^3$ [4] and 2.54 g/cm$^3$ [1], possibly due to impurities. Their samples "are not contaminated by any impurity *other than the capsule material*" [4], and contain a mixture B+PtB, a result of a reaction between boron and platinum capsule [1]; unfortunately, the concentration of Pt in the boron phase was not reported. Pt indeed enters the boron phase (affecting the color of the sample), rendering Pt capsules unsuitable [2]; it is far better to use inert pyrolytic BN capsules [8]. While chemically pure samples are dark grey [2], Zarechnaya et al. [1,4] report red samples. Contamination is a serious concern, in view of the known extreme sensitivity of boron to impurities [5,2]. Furthermore, high-pressure density seems to problematic. It is essential that static compression EoS parameters of incompressible phases ($K_0$=237 GPa [10]) be determined over the most hydrostatic conditions over a broad pressure regimes, much larger that 0-30 GPa, [1]; this was reported in Refs. [7,10], where markedly different result was obtained, more consistent with theory (Fig. 1).

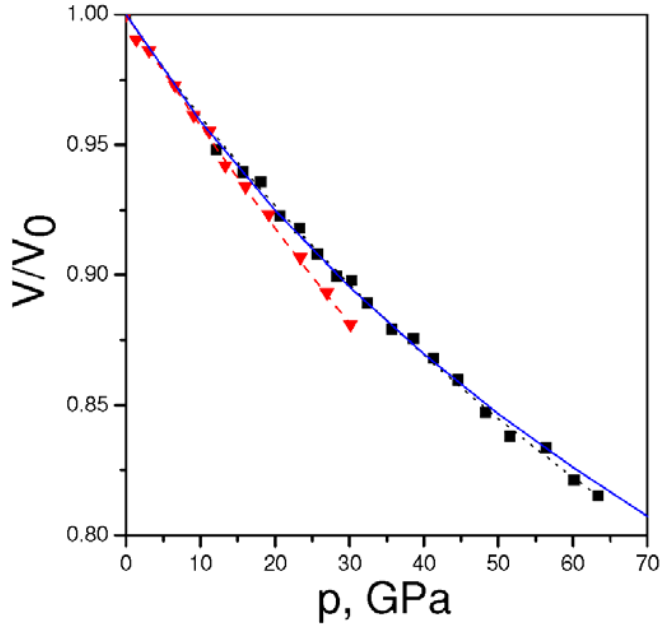

**FIG. 1. EoS of γ-B$_{28}$:** *red triangles* – experiment [1] (dashed line – our Vinet fit, $K_0$=235(6), $K_0$'=0.3(5)), *black squares* – experiment [10] (dotted line – Vinet fit, $K_0$=238(5) GPa, $K_0$'=2.7(3)), *solid line – ab initio* results [2] ($K_0$=222 GPa, $K_0$'=3.74 from a Vinet fit [2]).

Finally, we point out that the structure of γ-B$_{28}$ consists of B$_2$ and B$_{12}$ clusters in a NaCl-type arrangement [2] and can be represented as (B$_2$)$^{\delta+}$(B$_{12}$)$^{\delta-}$. Zarechnaya et al. [1] state that their results disagree with this model, meaning that either [1] reports another structure (which is not the case) or they find charge transfer $\delta$ to be zero (but no estimate of $\delta$ was made). Large covalent component of bonding [1] does not imply the absence of ionicity: mixed ionic-covalent (i.e. "polar covalent") bonding is extremely common. In γ-B$_{28}$, partial ionicity (our best estimate $\delta$~0.48 [2]) results from a difference in the electronic properties of B$_2$ and B$_{12}$ clusters, and affects physical properties, some of which are inexplicable within a purely covalent model [2]. E.g., LO-TO splitting stems from long-range electrostatic interactions between the atoms: "the LO and TO modes … are *nondegenerate* for *ionic* crystals…, whereas they are *degenerate* for *non-ionic (*homopolar) crystals" [15]. The presence of charge transfer can be concluded even from data of Zarechnaya et al. (Fig. 2 in [1]), as shown in Ref. [7].

Artem R. Oganov[1,2,*], Vladimir L. Solozhenko[3], Oleksandr O. Kurakevych[3], Carlo Gatti[4], Yanming Ma[5], Jiuhua Chen[6], Zhenxian Liu[7], Russell J. Hemley[7]

*Corresponding author

[1] *Department of Geosciences, Department of Physics and Astronomy, and New York Center for Computational Science, Stony Brook University, Stony Brook, NY 11794-2100, USA.*

[2] *Geology Department, Moscow State University, 119992 Moscow, Russia.*

[3] *LPMTM-CNRS, Université Paris Nord, Villetaneuse, F-93430, France*

[4] *CNR-ISTM Istituto di Scienze e Tecnologie Molecolari, via Golgi 19, 20133 Milano, Italy.*

[5] *National Lab of Superhard Materials, Jilin University, Changchun 130012, P. R. China.*

[6] *Center for the Study of Matters at Extreme Conditions and Department of Mechanical and Materials Engineering, Florida International University, Miami, FL 33199, USA*

[7] *Geophysical Laboratory, Carnegie Institution of Washington, Washington, DC 20015, USA*

1. Zarechnaya E.Y. et al., (2009). *Phys. Rev. Lett.* **102**, 185501.
2. Oganov A.R. et al., (2009). *Nature* **457**: 863–867.
3. Solozhenko V.L., Kurakevych O.O., Oganov A.R. (2008). *J. Superhard Mater.* **30**, 428-429.
4. Zarechnaya E.Y., et al. (2008). *Sci. Tech. Adv. Mat.* **9**, 044209.
5. Douglas, B.E., Ho, S.-M., Structure and Chemistry of Crystalline Solids (Springer, N.Y., 2006).
6. Wentorf, R. H., Jr. (1965). *Science* **147,** 49–50.
7. See EPAPS document No.
8. Solozhenko, V. L., Le Godec, Y. & Kurakevych, O. O. (2006). *C.R. Chimie* **9**, 1472–1475.
9. Amberger, E., Ploog, K. (1971). *J. Less-Common Metals* **23**, 21-31.
10. Le Godec Y., et al. (2009). Equation of state of orthorhombic boron, γ-$B_{28}$. *Solid State Comm.*, doi:10.1016/j.ssc.2009.05.025.
11. Irifune T., et al. (2003). *Nature* **421**, 599-600.
12. Mao W.L., et al. (2003). *Science* **302**, 425-427.
13. Li Q., et al. (2009). *Phys. Rev. Lett.* **102**, 175506.
14. Pan Z., et al. (2009). *Phys Rev. Lett.* **102**, 055503.
15. Elliott S.R. *The Physics and Chemistry of Solids*, page 247. John Wiley and Sons. (1998).

# EPAPS: Supplementary Information.

## Comment on 'Superhard Semiconducting Optically Transparent High Pressure Phase of Boron'

**I. Literature analysis.** Our assessment is that the paper of Zarechnaya et al. [E1] contains a large number of errors, and while our Comment (and subsequent sections of this EPAPS document) focus on problems with their original results, in this section we criticize their presentation of the knowledge in the field. In addition to the statements discussed in our Comment, Zarechnaya et al. [E1] make the following incorrect statements:

a. *"Several dozens of possible crystalline boron phases have been proposed in the literature, but most of the proved to be borides or to be stabilized by a small amount of impurities".*

We are aware of 16 reported phases (not "several dozens") [E2], and for most of them the experimental proof of chemical impurity is not existent.

b. *"Boron is known as a hard material (with the Vickers hardness reported as high as 25-30 GPa [24] for β-boron)"*

This gives the reader a distorted picture of the physical properties of boron, because the best estimates of the hardness of β-boron, as well as α-boron, are above 40 GPa (i.e. both are superhard) – 42 GPa for α-boron [E3] and 45 GPa for β-boron [E4].

c. *"We have found that samples [of γ-boron]… have a hardness of 58(5) GPa".*

The article by Zarechnaya et al. [E1] fails to mention that the hardness of γ-boron had been measured before in [E5], where a similar result, 50 GPa, was obtained. Ref. [E5] is cited in [E1], but no mention is given that the hardness was measured in [E5] – whereas the hardness was the only topic of ref. [E5].

d. *"the existence of ''tetragonal boron'' as a modification of pure elemental boron or as a boron-rich nitride or carbide has been a subject of controversy"* This statement ignores established facts and seems to mix up the two known and very different tetragonal phases: T-50 (proven to be a compound [E6]) and T-192 (pure elemental phase [E6]).

e. The claim (in the Abstract of [E1]) that γ-$B_{28}$ is the second hardest elemental material after crystalline diamond, ignores recent works [E7-E10].

f.  *"...the density of the B$_{28}$ phase is 2.54(1) g/cm$^3$, which is about 1% higher than densities of known (α-, β-, "tetragonal") boron modifications. This is not surprising, because only B$_{28}$ contains additional B atoms in an intericosahedral space (although in β-boron there are probably interstitial defect atoms"*.

It is well known that β-boron [E11-E13] and the tetragonal phase T-192 [E14] contain a very large number of intericosahedral atoms, and yet their densities are much lower than that of α-boron (which contains no intericosahedral atoms).

An important characteristic of a family of structures is the density difference - it affect high-pressure structural stability and calculation of such important characteristics as the Clapeyron slopes on the phase diagram. Contrary to [E1], γ-boron is not 1% denser than the other phases, as the calculation below shows.

**γ-boron:** Unit cell volume = 198.48 Å$^3$, 28 atoms/cell [E15, E1].

Volume per atom = 7.088 Å$^3$. Density = 2.533 g/cm$^3$.

**α-boron:** Unit cell volume = 263.5 Å$^3$, 12 atoms/cell [E16].

Volume per atom = 7.319 Å$^3$. Density = 2.453 g/cm$^3$. Density is 3.2% lower than for γ-boron.

**β-boron:** Unit cell volume = 2465.21 Å$^3$ [E7], 320 atoms/cell [E12, E13].

Volume per atom = 7.704 Å$^3$. Density = 2.330 g/cm$^3$. Density is 8.3% lower than for γ-boron.

**T-192 phase of boron:** Unit cell volume = 1456.95 Å$^3$, 192 atoms/cell [E14].

Volume per atom = 7.588 Å$^3$. Density = 2.366 g/cm$^3$. Density is 6.8% lower than for γ-boron.

**II. Details on charge transfer.** The crystal structure of γ-boron was solved in [E15] using *ab initio* evolutionary crystal structure prediction [E17] and experiment, and confirmed in [E18,E1, E19]. The structure has space group *Pnnm*, 28 atoms in the unit cell, and is a NaCl-type arrangement of B$_{12}$-icosahedra and B$_2$-pairs. The unusual characteristic of γ-B$_{28}$ is that there is significant charge transfer between the B$_2$ and B$_{12}$ clusters, as shown in [E15], and caused by their very different electronic properties. Structural formula of γ-boron can be represented as (B$_2$)$^{\delta+}$(B$_{12}$)$^{\delta-}$ with a mixed ionic-covalent bonding and non-zero charge transfer δ. Zarechnaya et al. [E1] objected to such a representation, claiming that the ionic component of chemical bonding is absent, but presented no evidence that δ=0.

What they showed instead (Fig. 1 of [E1]) is the accumulation of electron density between the atoms, which proves significant covalency of bonding, but does not imply that δ=0: electron density accumulations are well-known in such mixed ionic-covalent bonds as Si-O, where charge transfer and high degree of ionicity are undeniable [E20].

The other piece of evidence given by Zarechnaya et al. [E1] is that the projected electronic density of states shows that electrons belonging to $B_2$ pairs and $B_{12}$ icosahedra have the same range of energies (Fig. 2 of [E1]). This cannot be used as evidence against charge transfer [E21]. Careful analysis of these projected densities of states reveals, however, a more limited extent of hybridization and specific energy features.

Fig. E1 shows these projected densities of states (PDOS), which are similar to Fig. 2 of [E1]. PDOS depends on the localization procedure. The projection method we chose (Mulliken projection), imperfect like any other, possesses a useful property (unlike in [E1]) - the sum of PDOS equals the total DOS (note, however, that in Fig. E1 this is not obvious, since both the total DOS and the PDOS are reported per one atom, rather than per cell). We selected this kind of representation to facilitate comparison between the $B_2$ and $B_{12}$ sublattices, and for consistency with Ref. [E1]). Fig. E1 clearly shows that lowest-energy valence orbitals are dominated by the contributions from $B_{12}$ icosahedra, whereas the electrons around the Fermi level belong predominantly to the $B_2$ pairs. In these two important energy ranges the hybridization is very limited. Furthermore, the fact that lowest-energy electrons belong to the $B_{12}$ clusters, and highest-energy – to $B_2$ units, is consistent with the proposed [E1] direction of charge transfer: $B_2 \rightarrow B_{12}$.

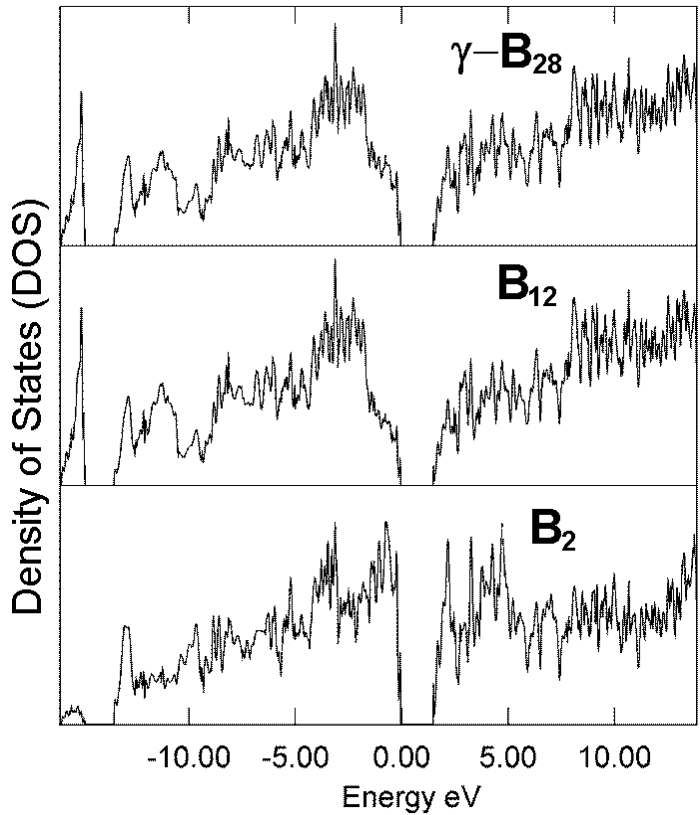

Fig. E1. Total electronic density of states of γ-boron and its projections (per atom) onto $B_{12}$ icosahedra and $B_2$ pairs.

This point can be made even clearer if one examines the spatial distribution of the electrons belonging to different energy ranges, so-called energy-decomposed electron density (EDED - Fig. E2). Indeed, coincidence of the energy range does not automatically imply hybridization: another necessary condition (not examined in [E1]) is the sharing of electron density in space. This analysis very clearly shows that lowest-energy electronic levels correspond to the $B_{12}$ icosahedra (energy range A), at intermediate energies there is indeed strong hybridization (range B), while the top of the valence band (range C) and the bottom of the conduction band (i.e. holes - range D) are dominated by the $B_2$ pairs. Strong localization of holes on the $B_2$ pairs is another evidence for charge transfer. An important ionic contribution to bonding was shown by us [E15] using several tools: (i) Bader analysis, (ii) spherical integration of electron density around the atoms, (iii) by considering electronic properties of the individual $B_{12}$ and $B_2$ sublattices, (iv) by examining lattice dynamics and proving the existence of significant long-range electrostatic interactions between the atoms (i.e. large Born charges, leading to LO-TO splitting and significant dielectric dispersion). Our preferred estimate of charge transfer, based on Bader analysis, is $\delta \sim 0.48$ [E15]. Zarechnaya et al. could have arrived at this conclusion, had their analyzed their own data carefully [E22].

Recently, we learned that γ-boron violates the Wade-Jemmis electron counting rules, and the only way to restore electronic balance is to assume charge transfer in the structure, i.e. partial ionicity [E23]. This also refutes the statement of Zarechnaya et al. [E1].

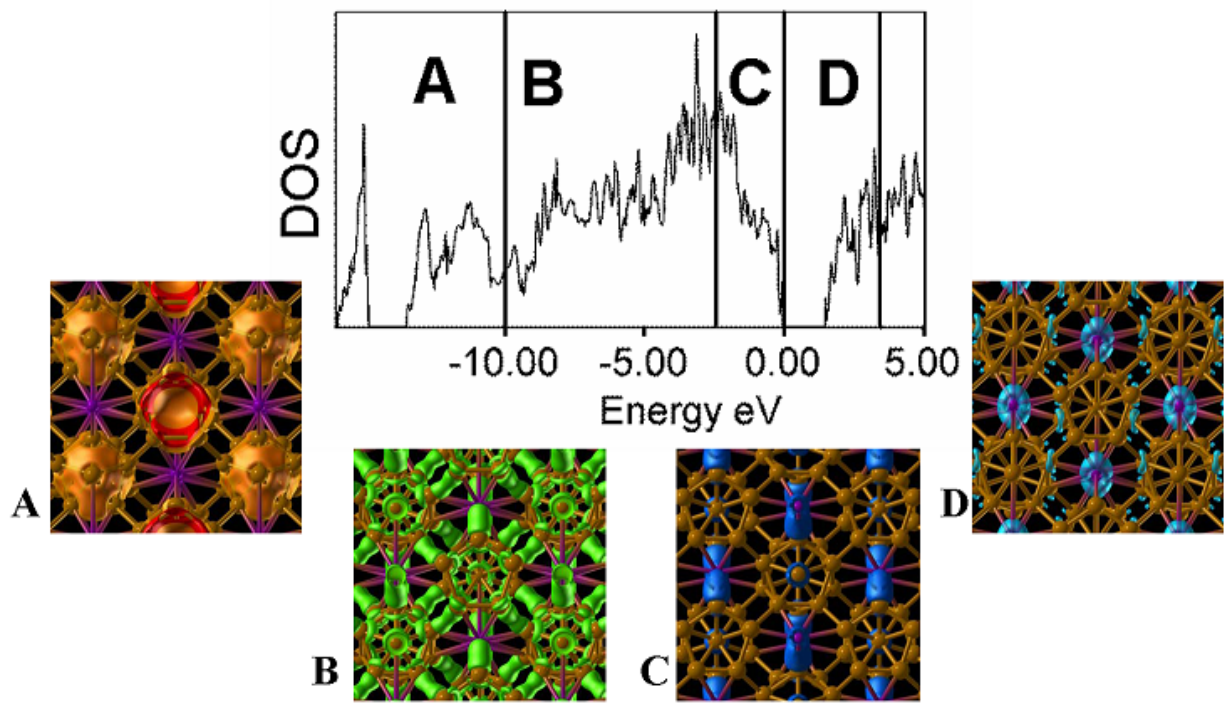

Fig. E2. Energy-decomposed electron density in γ-boron. The upper panel shows the total DOS and its separation into energy ranges A,B,C,D. Panels (A)-(D) show the corresponding electron density distributions.

### III. Details on the equation of state.

Zarechnaya et al. [E1] give parameters (Vo, Ko, Ko') of the equation of state, but do not specify the analytical form (e.g. Vinet, Murnaghan, Birch-Murnaghan, etc) to which these parameters were fitted. This renders their parameters meaningless, as we demonstrate on Fig. E3. As Fig. E3 shows, using the same set of parameters as reported in [E1] in conjuction with three commonly used analytical forms of the equation of state yield rather different results already at pressures where γ-boron is stable (i.e. at pressures below 89 GPa [E15])

a. Large discrepancies, up to 10 GPa, in the equation of state.

b. Spectacular divergence of the bulk modulus – already at pressures of up to 100 GPa, and these discrepancies rapidly increase on compression).

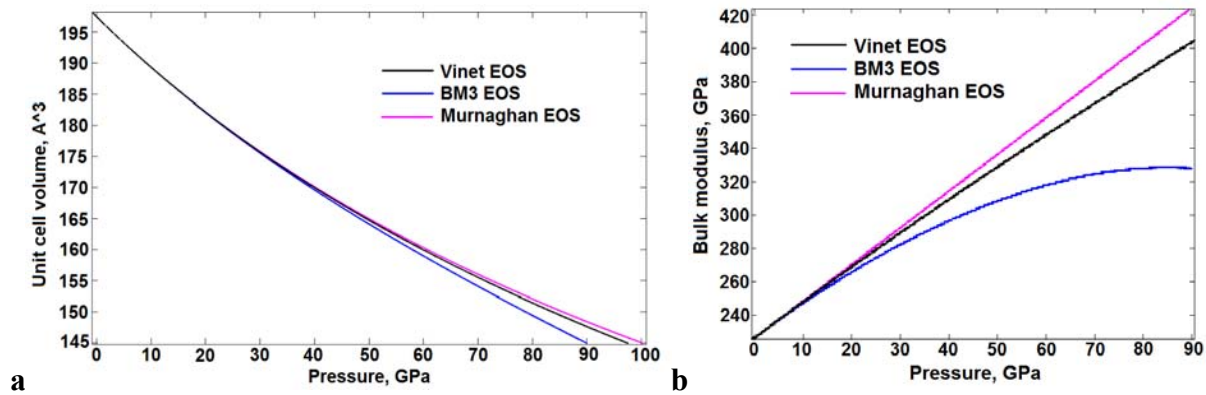

Fig. E3. (a) Equation of state and (b) bulk modulus as a function of pressure (derived from the equation of state) of γ-boron, calculated using parameters reported in [E1] and three commonly used analytical formulations of the equation of state (Vinet, 3$^{rd}$-order Birch-Murnaghan, and Murnaghan equations).

In view of this, we fitted parameters of the Vinet equation of state to theoretical data of Oganov et al. [E15], to experimental data of Zarechnaya et al. [E1] and to our experimental data [E24] obtained over a much wider pressure range. The results are:

|  | Vo, Å$^3$ | Ko, GPa | Ko' |
|---|---|---|---|
| Theory [E15] | 195.89 | 221.5(9) | 3.743(4) |
| Experiment up to 30 GPa [E1] | 197.44 | 235(6) | 0.3(5) |
| Experiment up to 65 GPa [E24] | 197.58 | 238(5) | 2.7(3) |

Experimental measurements of the equation of state of such an incompressible phase only to ~30 GPa (as done in Ref. [E1]) are inadequate for constraining the Ko' parameter, which probably explains the large discrepancy with the theoretical value of this parameter. At the same time, the experimental equation of state obtained in quasihydrostatic conditions to 65 GPa [E24] matches *ab initio* calculations [E15, E1], including the Ko' parameter, much better.

Artem R. Oganov[1,2,*], Vladimir L. Solozhenko[3], Oleksandr O. Kurakevych[3], Carlo Gatti[4], Yanming Ma[5], Jiuhua Chen[6], Zhenxian Liu[7], Russell J. Hemley[7]

*Corresponding author

[1] *Stony Brook University, Stony Brook, NY 11794-2100, USA.*

[2] *Moscow State University, 119992 Moscow, Russia.*

[3] *LPMTM-CNRS, Université Paris Nord, Villetaneuse, F-93430, France*

[4] *CNR-ISTM Istituto di Scienze e Tecnologie Molecolari, via Golgi 19, 20133 Milano, Italy.*

[5] *Jilin University, Changchun 130012, P. R. China.*

[6] *Florida International University, Miami, FL 33199, USA*

[7] *Carnegie Institution of Washington, Washington, DC 20015, USA*

**References.**

E1. Zarechnaya E.Y., et al. (2009). *Phys. Rev. Lett.* **102**, 185501. Submitted 16 January 2009, published 7 May 2009.
E2. Douglas, B.E., Ho, S.-M., Structure and Chemistry of Crystalline Solids (Springer, N.Y., 2006).
E3. Amberger, E. and Stumpf, W., Boron, *Gmelin Handbook of Inorganic Chemistry*, Berlin: Springer_Verlag, 1981, pp. 112–238.
E4. Gabunia D., et al. (2004). *J. Solid St. Chem.* **177**, 600-604.
E5. Solozhenko V.L., Kurakevych O.O., Oganov A.R. (2008). *J. Superhard Mater.* **30**, 428-429.
E6. Amberger, E., Ploog, K. (1971). *J. Less-Common Metals* **23**, 21-31.
E7. Irifune T., et al. (2003). *Nature* **421**, 599-600.
E8. Mao W.L., et al. (2003). *Science* **302**, 425-427.
E9. Li Q., et al. (2009). *Phys. Rev. Lett.* **102**, 175506.
E10. Pan Z., et al. (2009). *Phys Rev Lett.* **102**, 055503.
E11. Slack G.A., et al. (1988). *J. Sol. State Chem.* **76**, 52-63.
E12. Widom M., Mikhalkovic M. (2008). *Phys. Rev.* **B77**, 064113.
E13. Ogitsu T., et al. (2009). *J. Am. Chem. Soc.* **131**, 1903-1909.
E14. Vlasse, M., Naslain, R., Kasper, J.S., Ploog, K. (1979). *J. Sol. State Chem.* **28**, 289-301.
E15. Oganov A.R., et al. (2009). *Nature* **457**, 863–867. Submitted 27 January 2007, Published online 28 January 2009.
E16. Will, G., Kiefer, B. (2001). *Z. Anorg. Allg. Chem.* **627**, 2100-2104.
E17. Oganov A.R., Glass C.W. (2006). *J. Chem. Phys.* **124**, 244704.
E18. Zarechnaya E.Y., et al. (2008). *Sci. Tech. Adv. Mat.* **9**, 044209. Submitted 3 November 2008, published online 28 January 2009.

E19. Authors of [E1,E12] had received from us manuscript [E10] on 9 December 2006. Neither this, nor the earlier publication of our results in the form of an abstract [C. Gatti, et al. (2008). *Acta Cryst.* A64, C70 (August 2008)] was acknowledged by Zarechnaya et al. [E1,E12].
E20. Tsirelson V.G., et al. (1990). *Phys. Chem. Min.* **17**, 275-292.
E21. According to the Manual of VASP code (used in [E1]), for the projected densities of states "results are qualitative — i.e. there is no unambiguous way to determine the location of an electron. With the current implementation, it is for instance hardly possible to determine charge transfer." (http://cms.mpi.univie.ac.at/vasp/vasp.pdf, section 6.32).
E22. We have requested technical information (input and output files) from Zarechnaya et al., but did not receive any response.
E23. Rulis P., Wang L., Ching W.Y. (2009). *Phys. Stat. Sol.*, doi:10.1002/pssr.200903089
E24. Le Godec Y., et al. (2009). Equation of state of orthorhombic boron, γ-$B_{28}$. *Solid State Comm.*, doi:10.1016/j.ssc.2009.05.025.

**Added note:** In their 2010-paper, Dubrovinskaia et al. (Zarechnaya E., et al. (2010). Pressure-induced isostructural phase transformation in γ-$B_{28}$. *Phys. Rev.* **B82**, 184111) used partial ionicity (i.e. polarity) of bonding in γ-$B_{28}$ for explaining anomalies in its compressive behavior. Thus, in their later work Dubrovinskaia et al. have acknowledged that their initial judgment on the chemical bonding in γ-$B_{28}$ was incorrect.